\documentclass[11pt,a4paper,oneside]{article}

\pdfoutput=1

\usepackage{amsfonts}
\usepackage{amsmath}
\usepackage{amssymb}
\usepackage{float}
\usepackage[a4paper]{geometry}
\usepackage{graphicx}
\usepackage{a4wide}
\usepackage{ifthen}
\usepackage{fancyhdr}
\usepackage{color}
\usepackage{colortbl}
\usepackage{multirow}
\usepackage{eso-pic}

\usepackage{graphicx}
\usepackage{hyperref}

\begin{document}

\title{Empirical forward price distribution from Bitcoin option prices}
\author{Nikolai Zaitsev\footnote{Director of UXTA.io and INNOVAEST.ORG. Correspondent e-mail: nikzaitsev@yahoo.co.uk}}
\date{12 Jan 2019}

\maketitle

\begin{abstract}
Report presents analysis of empirical distribution of future returns of bitcoin (BTC) from BTUSD inverse option prices. Logistic pdf is chosen as underlying distribution to fit option prices. The result is satisfactory and suggests that these prices can be described with just three or even one parameter. Fitted Logistic pdf matches forward price movements upto a scaling factor. Nevertheless, this observation stands alone and does not allow stochastic description of underlying prices with logistic pdf in similar fashion as it is done within Black-Scholes modelling framework. Put-call parity relationship is derived connecting prices of vanilla inverse options and futures.
\end{abstract}

\section{Introduction}

\subsection{Crypto market and Bitcoin (BTC) options}
\label{sec:crypto_market}

Crypto coins market is quite new, however, due to the recently inflated bubble peaked to the size of \$800 billion and then imploded to the level of \$100 billion it attracted great attention from investors. Barely noticed 4-5 years ago it is considered today as an asset class attractive for investments. Despite such attention, it still lacks official recognition by the governmental authorities.

Bitcoin (BTC) is the first and the most popular crypto coin. It consistently leads all other coins by the market capitalization and trading volume. It is traded against fiat currencies, such as dollar (USD), euro (EUR), sterling pound (GBP) and many others across multiple exchanges (for example, see the list at~\cite{bitcoin.org} or~\cite{coinmarketcap.com}. BTC is attractive due to various reasons, where a possibility for direct transactions between two counterparts (P2P transactions) and strictly controlled inflation of monetary base are the major one. Since Bitcoin regulation and legal protection are still in their infancy it is vulnerable to price manipulation either from media or from traders with large financial resources. As result, BTC exchange rate is volatile with sharp jump-like movements. Such dynamics creates demand from crypto investors for hedging their risks and for pure speculations.

This demand stimulates foundation of derivative exchanges supporting public trading in options, futures, perpetuals, CFD's, swaps and other products. For example, BTCUSD Futures are traded also at CBOE (Chicago Board Option Exchange) within regulated environment. This report focuses on options. So far only three option exchanges are known:

\begin{itemize}
\item Deribit ( \url{https://deribit.com} )
\item Quedex ( \url{https://quedex.net} )
\item LedgerX ( \url{https://ledgerx.com} )
\end{itemize}

They all support trading of options and futures, however they differ in data policy. Deribit exchange has open data policy and allows very simple download of snapshot of option and futures prices in the trading book. Information of trading book of Quedex requires registration and some additional programming effort, however, it is also open. LedgerX is an OTC platform targeting large investors. Their data is not publicly available. This report uses Deribit data only.

Aim of this report is extraction of forward-looking price distribution from analysis of option prices. Its purpose is to give reader an impression about market sentiments and provide tools for evaluation of risks present in the market. To give more faith to this analysis the report gives a brief description about what options do. 

This report is based on the analysis of empirically observed market data and tries to follow model free path. The results are fit to logistic distribution function, which is not commonly used in financial mathematics. The resulting fit is surprisingly good. Solution of this puzzle is left beyond this article, which aims only to report the observed fact. 

\section{Contract definitions}
\label{sec:contract_definitions}

Option is a contract allowing buyer the right to receive payments according to the predefined payout formula. This formula is calculated at the \textit{settlement} which occurs at \textit{Maturity} date. Price of \textit{Underlying} at \textit{Maturity} is input to the formula applied. This report discusses \textit{vanilla European Put} and \textit{Call} options and BTCUSD as \textit{Underlying}. Deribit exchange is marketing contracts settled at specified maturities in accord to its \textit{Type} (\textit{Call} or \textit{Put}) and \textit{Strike}. Before maturity, the contracts may change hands via trading between clients of exchange. Number of contracts circulating around is named \textit{Open interest} and is specified per type and strike. \textit{Open interest} is not included into analysis presented in this report.

Strike level is expressed in USD paid for 1 BTC (i.e. \textit{BTCUSD} exchange rate). Option price is settled "physically" i.e. the account is always debited or credited in units of BTC. This happens since many crypto exchanges operate outside of any regulations.

BTC is not exchanged for USD at Deribit. Instead the exchange uses external quote. Since BTC price is subject to large jumps, Deribit uses an average of rates quoted at few liquid exchanges, such as Bitstamp, GDAX (Coinbase), Gemini, Itbit, Kraken and Bitfinex~\footnote{during observation period presented in this paper, Deribit did not include Bitfinex into the average}. The averaging algorithm is explaned as follows:

\begin{itemize}
\item 6 online: we leave the highest and lowest and each exchange accounts for 25\%
\item 5 online: we leave the highest and lowest and each exchange accounts for 33.33\%
\item 4 online: we leave the highest and lowest and each exchange accounts for 50\%
\item 3 online: each exchange accounts for 33.33\%
\item 2 online: each exchange accounts for 50\%
\item 1 online: exchange represents the index
\end{itemize}

\section{Mathematical definitions}
\label{sec:math_definitions}

At Deribit call (put) is defined as a contract where buyer has a right to buy (sell) the underlying, $x$, at fixed price level called strike, $K$. At the time of maturity ($t=T$) buyer may execute this right only if price of underlying, $x_T$, is above (below) strike level for call (put). Options deliver bitcoins (BTC) while underlying and strike are expressed in terms of underlying exchange rate. The payout of call is defined as:

\begin{equation} \label{eq:call_def}
	C(x_T,K)=\frac{max⁡(0,x_T-K)}{x_T} =\frac{(x_T-K)^+}{x_T}
\end{equation}
Similarly, payout of put is defined as:
\begin{equation} \label{eq:put_def}
	P(x_T,K)=\frac{max⁡(0,K-x_T)}{x_T} =\frac{(K-x_T)^+}{x_T}
\end{equation}

On the other side of the contract, call (put) seller (also underwriter) is obliged to sell (buy) the \textit{Underlying} at price defined by the \textit{Strike}. Options and futures are settled in bitcoins. Because of that we see two consequences: 
\begin{itemize}
\item non-zero part of option payout is non-linear.
\item No fiat currency, such as USD, is involved in the exchange of funds between traders. Hence, USD account is excluded from further consideration. 
\end{itemize}

\section{Properties of option prices}
\label{sec:properties_of_option_prices}

From the definition of payout one can derive one useful relationship named put-call parity.
Commonly used notations are:
\begin{itemize}
\item At-the-money (ATM) level. It is the level where the future x price is equal to strike, $K$.
\item In-the-money (ITM) option is call (put) option where price of underlying is significantly larger (less) than strike. In this case, 
there is very high correlation between movements of option price and price of underlying.
\item Out-the-money (OTM) option is a call (put) option where price of underlying is significantly less (larger) than Strike. In this case, the correlation between movements of option price and price of underlying is very low.
\end{itemize}

\subsection{Put-call parity}
\label{sec:put_call_parity}

It is easy to infer that under any circumstances at Maturity we will have the following:
\begin{align} \label{eq:put_call_parity_at_t}
\begin{split}
C(x_T,K,T)- P(x_T,K,T)=\frac{(x_T-K)^+}{x_T} - \frac{(K-x_T)^+}{x_T}\\
C(x_T,K,T)- P(x_T,K,T)=\frac{x_T-K}{x_T}
\end{split}
\end{align}
This relationship is called “put-call parity” and it is also valid any moment before maturity. This fact can be rewritten in terms of expectations. Since the future price outcome is not known one can apply expectation operator $E_t [.]$ to both sides:
\begin{align} \label{eq:put_call_parity_any}
\begin{split}
E_t[C(x_T,K,T)]- E_t[P(x_T,K,T)]=E_t[\frac{{x_T-K}}{x_T}]\\
=1-K \cdot E_t [\frac{1}{x_T}]
\end{split}
\end{align}
Eq.~\ref{eq:put_call_parity_at_t} and Eq.~\ref{eq:put_call_parity_any} become equivalent at $t=T$. Further we use that $E_t [C(x,K,T)]=C(x,K,t)$:
\begin{equation} \label{eq:put_call_parity_any2}
C(x,K,t)- P(x,K,t)=1 - K \cdot E_t[\frac{1}{x_T}]
\end{equation}
BTC has no discount mechanism due to no embedded inflation. To calculate expectation of inverted value we have to define “inverse” futures~\cite{ABragin}, non-linear products, traded at the same exchange~\footnote{It is derived, by looking from bitcoin perspective, where $y_t=\frac{1}{x_t}$. Without discount we have $F_t^{(y)} = E_t[y_T] = y_t$. Hence, $F_t = 1/F_t^{(y)}$}:
\begin{equation} \label{eq:futures_def}
\frac{1}{F_t} =E_t [\frac{1}{x_T} ]	
\end{equation} 
Finally, substitute, this expression into Eq.~\ref{eq:put_call_parity_any2}:
\begin{equation} \label{eq:put_call_parity_final}
C(x_t,K,t)- P(x_t,K,t) = 1 - \frac{K}{F_t}
\end{equation}
This relationship does not depend on evolution of underlying. Eq.~\ref{eq:put_call_parity_final} is one of results of this note written for 'inverted' options and futures on BTC. Note, the unusual for put-call parity relationship between Long Call, Short Put and inverse futures. That is because of the physical settlement feature of these contracts. 

\subsection{Expected future distribution of BTC}
\label{sec:expected_future_btc}

Beliefs of people about different outcome of $x_T$ given current observation of $x_t$ is reflected through option prices they are quoting and subsequently trade. To see that we have to define a function of probability to observe price of underlying of $x_T$ at maturity given the initial value of $x_t$: $f(x_T |x_t )$. The expectation operator used above can be rewritten as (use call as an example):
\begin{align} \label{eq:option_pricing}
\begin{split}
E_t [C(x_T,K,T)]=C_t (x_t,K,t)=\int_{-\infty}^{+\infty} (1-K/x_T )^+ \cdot f(x_T |x_t ) \cdot dx_T\\
= \int_K^{+\infty} (1-K/x_T )\cdot f(x_T |x_t )\cdot dx_T  	
\end{split}
\end{align}

Function $f(x_T |x_t )$ follows properties of probability function and can be interpreted as a probability to find a person who 'believes into' $x_T$ given spot value $x_t$. People within marketplace negotiate between each other in accordance to their view of the future, where the option price results in the recorded trade.
Hence, option price as a function of strike, $K$, and of spot price, $x_t$, contains information about distribution of underlying which is believed by traders to be at $t=T$. There is simple model independent approach to extract this information. Note, that Eq.~\ref{eq:option_pricing} is written without making any specific assumptions about shape of $f(x)$. By taking first derivative of call price function with respect to strike we find that:
\begin{equation} \label{eq:cds_from_option}
	\frac{dC_t (x_t,K)}{dK}=\frac{1}{x_{T}} \cdot (1-CDF(x_{T}=K|x_{t}))  	
\end{equation}
By taking second derivative one can observe the result from Breeden and Litzenberger \cite{breeden}, which is distribution of future BTC price:
\begin{equation} \label{eq:pdf_from_option}
	\frac{d^2 C_t (x_t,K)}{dK^2}=\frac{f(x_T=K│x_t )}{x_T}
\end{equation}
Similar expression is true for puts.

\subsection{Logistic distribution}
\label{sec:logistic_distribution}

So far, there was no assumption about the shape of $f(x)$. Log-Normal distribution is used more frequently as it lays the basis under modern option pricing theory. Original derivation of Black-Scholes formula uses arbitrage-free arguments. This model introduces volatility as a parameter, which is widely used in the investments and trading environment. At some point, \cite{derman}, it was found that volatility implied from option prices exhibits dependency on strike and is asymmetric. This feature is called "volatility smile". Such behavior indicates that the terminal p.d.f. (or $f(x)$) does not follow assumed Log-Normal distribution.
In attempt to resolve this issue empirically, let us look onto CDF properties, where:

\begin{align} \label{eq:cdf_property}
\begin{split}
\lim_{x \to  -\infty}⁡ CDF(u)=0\\ 
\lim_{x \to +\infty} CDF(u)=1	
\end{split}
\end{align}
One may notice that CDF-function resembles sigmoid function, which is:
\begin{equation} \label{eq:cdf_sigmoid}
CDF(x,m,s) =\frac{1}{1+e^{-\frac{x-m}{s}}}
\end{equation}
The respective p.d.f. is referred as Logistic distribution, which is:
\begin{equation} \label{eq:pdf_logistic}
PDF(x,m,s)= f(x,m,s)=\frac{e^{-\frac{x-m}{s}}}{s \cdot (1+e^{-\frac{x-m}{s}})^2 }
\end{equation}
By recalling the relationship between Eq.~\ref{eq:cds_from_option} and Eq.~\ref{eq:pdf_from_option} one may naturally think to use sigmoid to model $CDF$-function. The advantage of using sigmoid as $CDF$ is that all its "siblings" (i.e. $CDF(x)$, $PDF(x)$) have close-form. An integral of sigmoid ($IS(x,m,s)$) has closed-form as well:
\begin{equation} \label{eq:integral_sgimoid}
IS(x,m,s)=\log⁡(1+e^\frac{x-m}{s}) \cdot s
\end{equation}
Here we may interpret parameter $m$ as $ATM$-level and $s$ as a volatility.
Considering these similarities, further we use $IS(x)$ to model Put price, i.e. $IS(x) \to P_{th}(x)$, as a function of strike. Additional parameter $a$ is added: 
\begin{equation} \label{eq:put_as_is}
P_{th}(x,m,s)=\log⁡(1+e^\frac{x-m}{s}) \cdot s \cdot a
\end{equation}
This parameter is responsible for normalization of CDF, because: $CDF(+\infty) = a$. In our analysis all three parameters are left free. Model with two fixed parameters is also checked. Let us apply this guess to the data.

\section{Analysis}
\label{seq:analysis}

\subsection{Data}
\label{seq:data}

Data used in this analysis, are downloaded from Deribit and are collected during a day at 5-minutes intervals. It has open REST feed, which delivers snapshot of limit order book for options and futures quoted over all marketed Maturities. As of today, Deribit quotes two futures with maturities in Dec18, and Mar19, while there are two option maturities in Dec18, one in Jan19, Mar19 and Jun19, i.e. five option maturities in total.

Data were collected from 10-Dec-2018 to 16-Dec-2018.

\subsection{At-the-money level}
\label{seq:at_the_money}

ATM level is equal to forward value of underlying. As a number this can be measured in three ways:
\begin{itemize}
\item calculate forward value directly through discounted cash flows method (fair valuation). 
\item use observed futures prices at the same maturity as of options under question. Deribit quotes less futures maturities than that for options.
\item by using call-put parity one can build synthetic futures prices as function of strike and find the zero price via linear regression. By put-call parity, it is futures price.
\end{itemize}

\begin{figure}[h]
\centering
\includegraphics[width=1.1\textwidth]{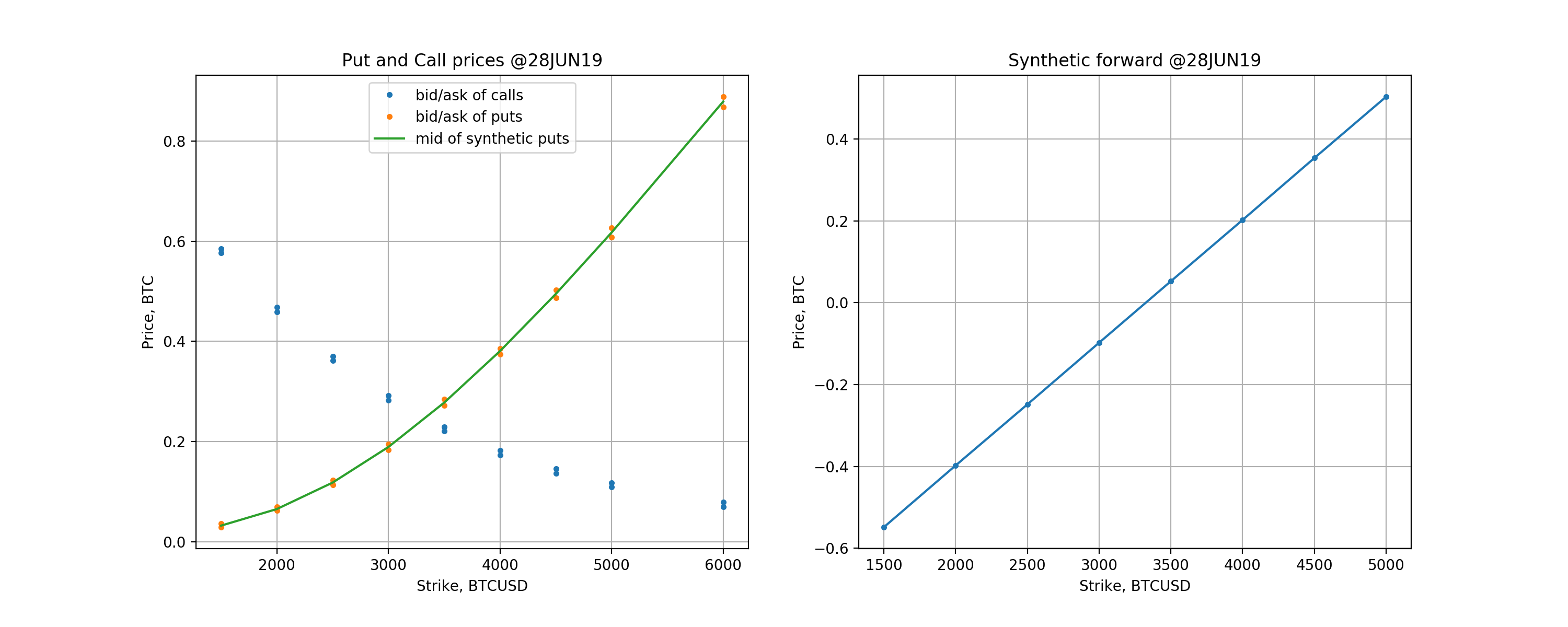}
\caption{Left: Call (blue) and Put (orange) bid and ask prices as function of strike. Green line shows synthetic put mid prices, $P_c(x)$. Right: Price of synthetic forward as Long Call Short Put price as a function of strike. Options of maturity at 29-Mar-19 are used. Snapshot is taken at 2018-Dec-11 04h:10m}
\label{fig:call_put_prices}
\end{figure}

The latter method is used to imply ATM level ($E_t [1/x_T]^{-1}$) for all option maturities. See Fig.~\ref{fig:call_put_prices}, Right plot for illustration. Further we use ATM-level and not futures for two reasons:
\begin{itemize}
\item it is found empirically, through option prices, therefore, it is consistent with chosen set of option prices
\item there are more option maturities than futures maturities
\end{itemize}

\subsection{Filter option prices}
\label{seq:filter_prices}

Both, call and put, prices contain the same information about future price of underlying. Therefore, call prices are converted to put prices by using call-put parity:
\begin{equation} \label{eq:call_to_put}
P_c (x)=C(x)-1+\frac{K}{ATM}
\end{equation}
See calls and puts prices on Fig.~\ref{fig:call_put_prices}, Left plot for an illustration of the method. 

As result of data cleansing, we have 4 points per strike, which are averaged into "combined put prices". Resulting option prices require certain filtering. Although in general we find smooth data, at night (all crypto exchanges operate on 24x7 basis) or during extreme price movements quotes are moving away or even disappear. Only option prices with both ask (sell) and bid (buy) quotes are used.

\subsection{Fitting put prices to integral sigmoid}
\label{sec:fit_prices_to_is}

\begin{figure}[h]
\centering
\includegraphics[width=1.0\textwidth]{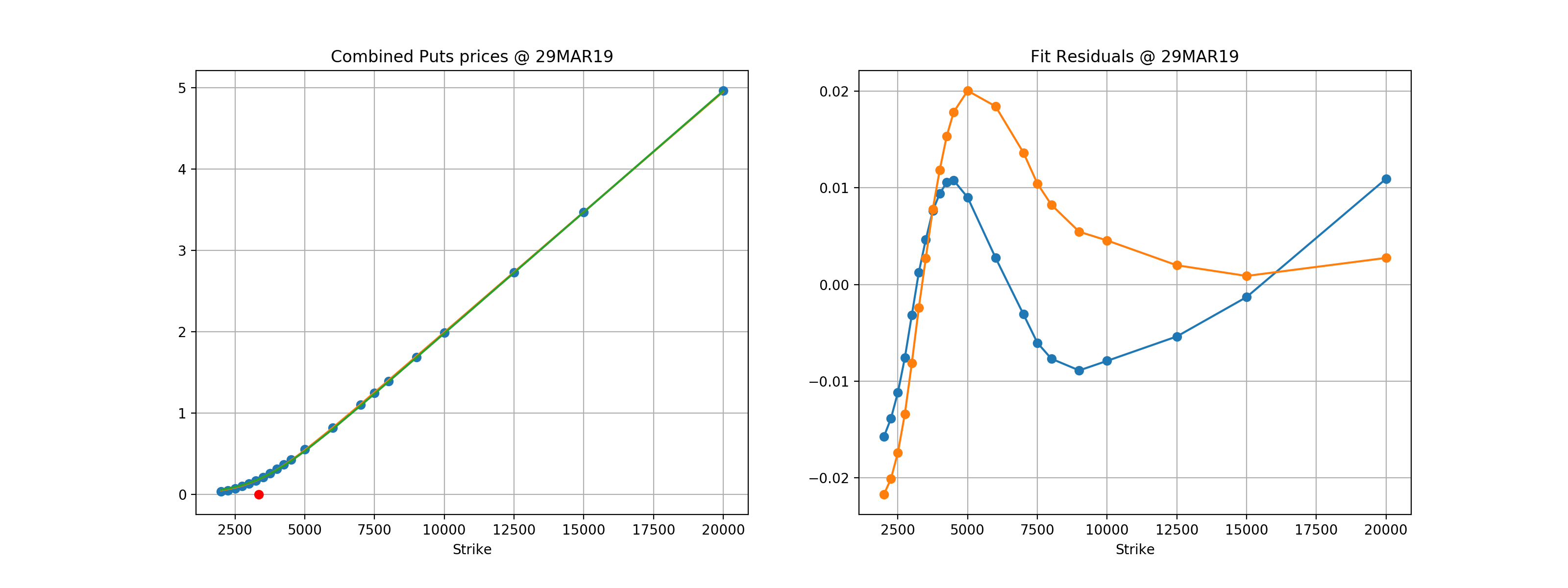}
\includegraphics[width=1.0\textwidth]{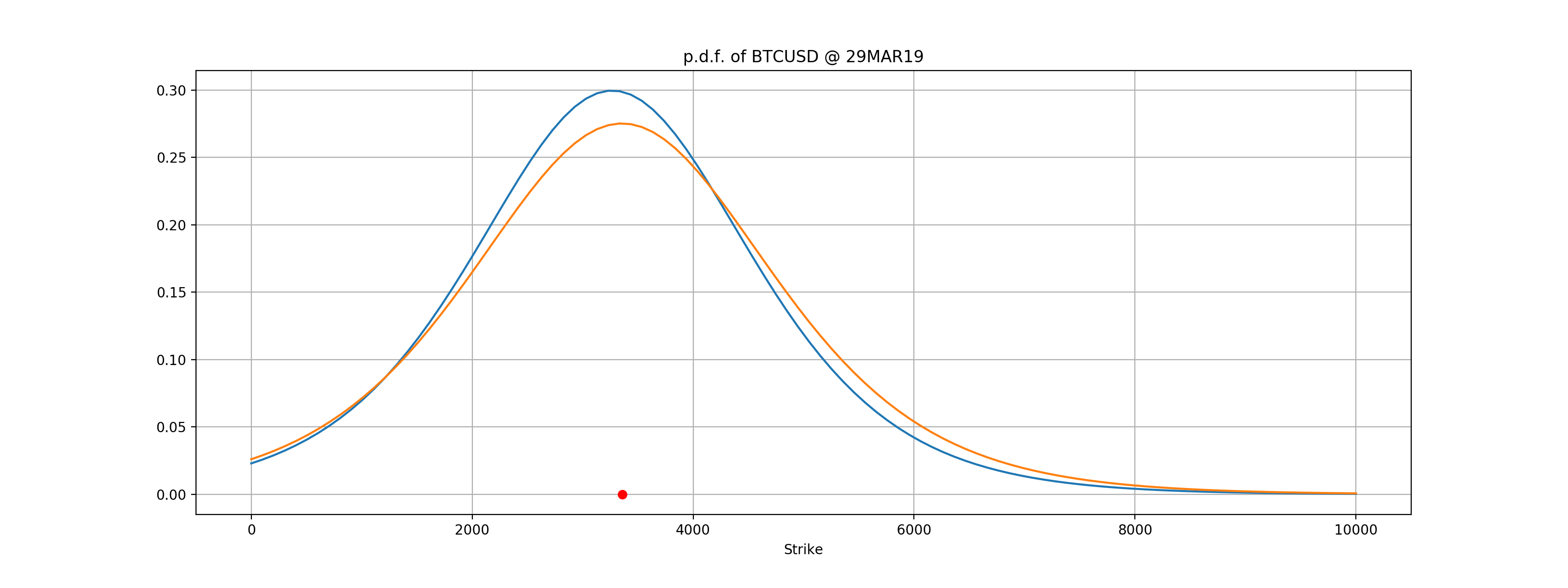}
\caption{Top-Left: Fit (line) of integrated sigmoid to combined put prices (dots). Top-Right: Fit Residuals with respect to mid-prices for 3-parameter ($msa$) fit (Blue) and $s$-parameter fit (Orange). Bottom: Respective p.d.f of forward BTCUSD for  3-parameter ($msa$) fit (Blue) and $s$-parameter fit (Orange). Red point shows ATM-level. Options of maturity at 29-Mar-19 are used. Snapshot is taken at 2018-Dec-11 04h:10m}
\label{fig:fit_is_pdf_mar19}
\end{figure}

\begin{figure}[h]
\centering
\includegraphics[width=1.0\textwidth]{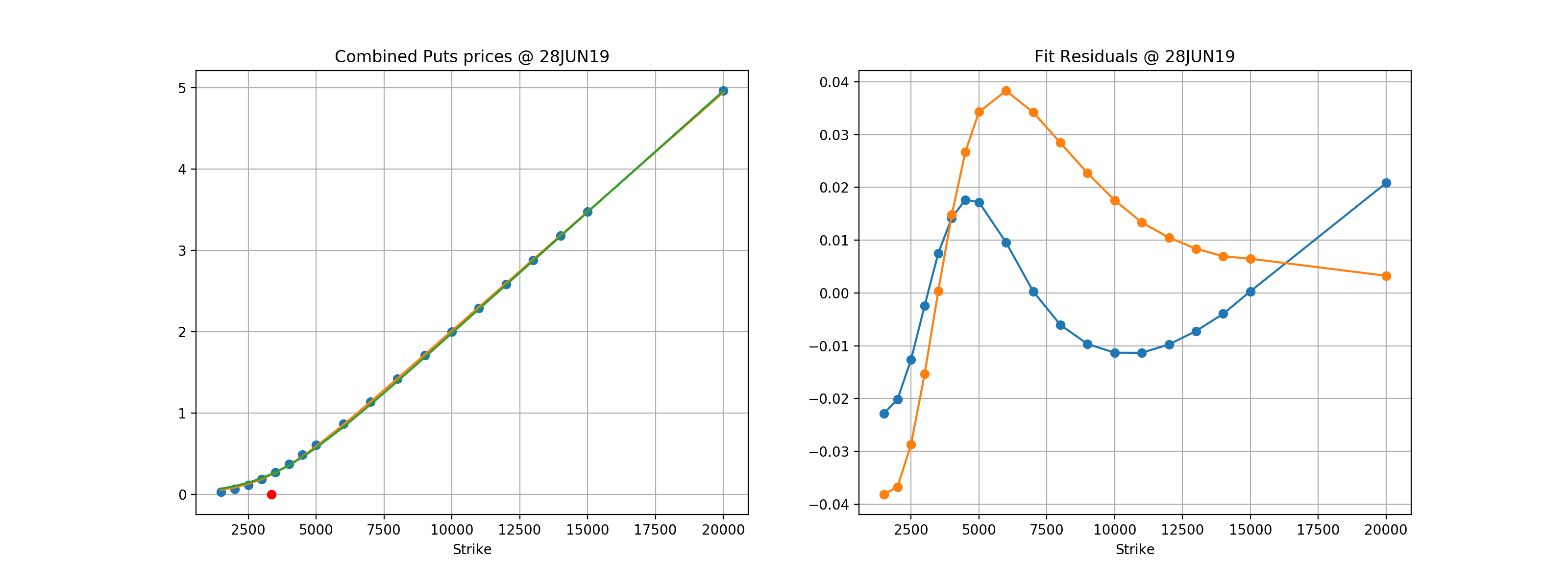}
\includegraphics[width=1.0\textwidth]{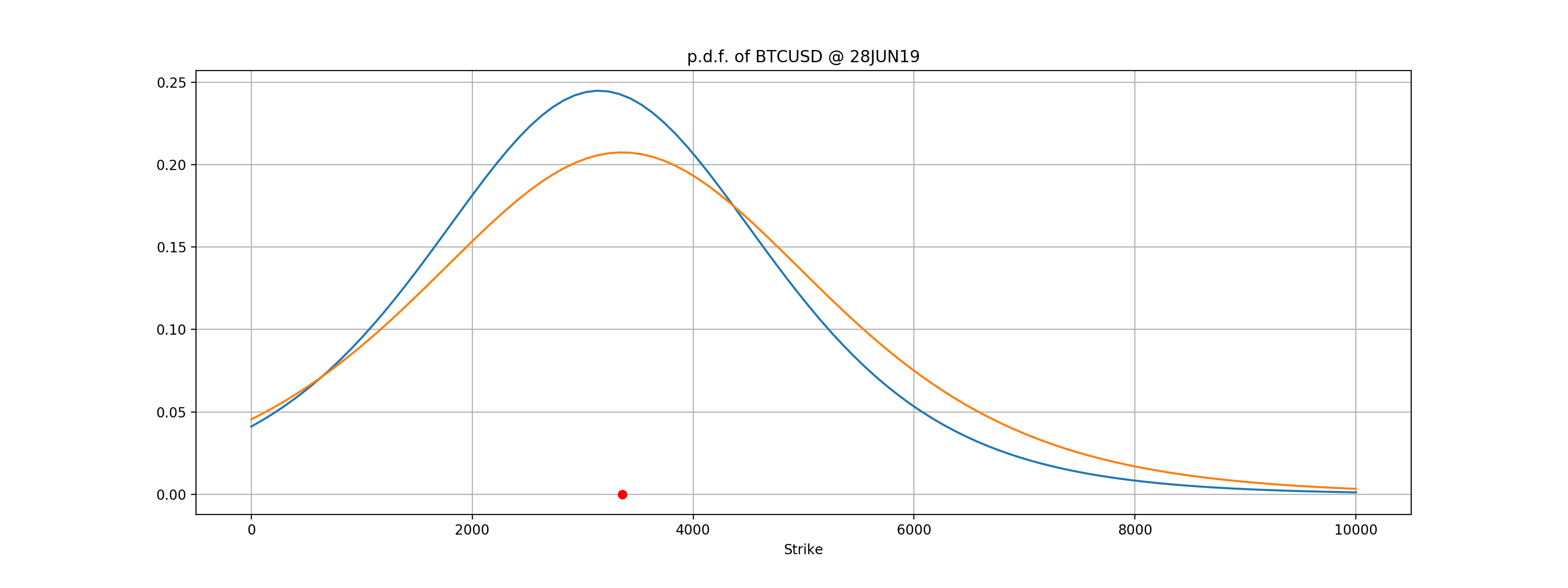}
\caption{Top-Left: Fit (line) of integrated sigmoid to combined put prices (dots). Top-Right: Fit Residuals with respect to mid-prices for 3-parameter ($sma$) fit (Blue) and $s$-parameter fit (Orange). Bottom: Respective p.d.f of forward BTCUSD for  3-parameter ($sma$) fit (Blue) and $s$-parameter fit (Orange). Red point shows ATM-level. Options of maturity at 28-Jun-19 are used. Snapshot is taken at 2018-Dec-11 04h:10m}
\label{fig:fit_is_pdf_jun19}
\end{figure}

In the next step, integrated sigmoid from Eq.~\ref{eq:put_as_is} is fitted to the combined put prices by using Least Squares method (see Fig.~\ref{fig:fit_is_pdf_mar19} and Fig.~\ref{fig:fit_is_pdf_jun19}). Two functions were applied. First, is when all three parameters, $m$, $s$, $a$, are free. Second, is when only $s$-parameter is free and others are fixed as $a=1$ and $m=ATM$. First method has smaller residuals, but the second one fits almost within bid-ask. This indicates that we can reduce option description to just two parameters, $ATM$ and $s$, one of which is 'almost' observable and can be replaced with futures price.

Prices fit integrated sigmoid very well. Average residual between data and theoretical points is of the order or better than the average bid-ask spread of option quotes. This means that all option prices per each maturity can be parameterized with just three parameters, m, s and a or even with only one parameter. For details see Table below.

Span of p.d.f. into range of negative strikes may be interpreted as a possibility for BTC to "default". Although BTC cannot default, this term might illustrate the situation where BTC price may become (near) zero. For example, implied probability of default ($IPD$) for distributions shown in Fig.~\ref{fig:fit_is_pdf_mar19} and Fig.~\ref{fig:fit_is_pdf_jun19}, are 0.87\% and 2.50\% respectively. Summary of findings for this snapshot is given in the Table~\ref{tables:ipd}:

\begin{table}
\begin{center}
\small{
\begin{tabular}{|r|rrrrrrr|}
\hline
МMaturity & $IPD$, \% & $m$ & $s$ & $a$ & $Res (\times 1000)$ & $Spr (\times 1000)$ & $s (single)$\\
\hline
28-Dec-18	& 0.00 & 3.52 & 0.31 & 1.00 & 1.12 & 5.69 & 0.45 \\
25-Jan-19	& 0.13 & 3.38 & 0.51 & 0.95 & 2.04 & 5.88 & 0.69 \\
29-Mar-19	& 0.87 & 3.22 & 0.68 & 0.95 & 3.98 & 6.43 & 0.90 \\
28-Jun-19 & 2.50 & 2.95 & 0.81 & 0.92 & 4.72 & 10.36 & 1.19 \\
\hline
\end{tabular}
\caption{Implied PD, parameters, fit residual and average bid-ask spread are shown for different maturities. For fit stability, underlying prices were divided by 1000. $m$-parameter is 1000 smaller as it should be. $s$(single) is fit result of 1 parameter model.}
\label{tables:ipd}
}
\end{center}
\end{table}

\begin{figure}[h]
\centering
\includegraphics[width=0.7\textwidth]{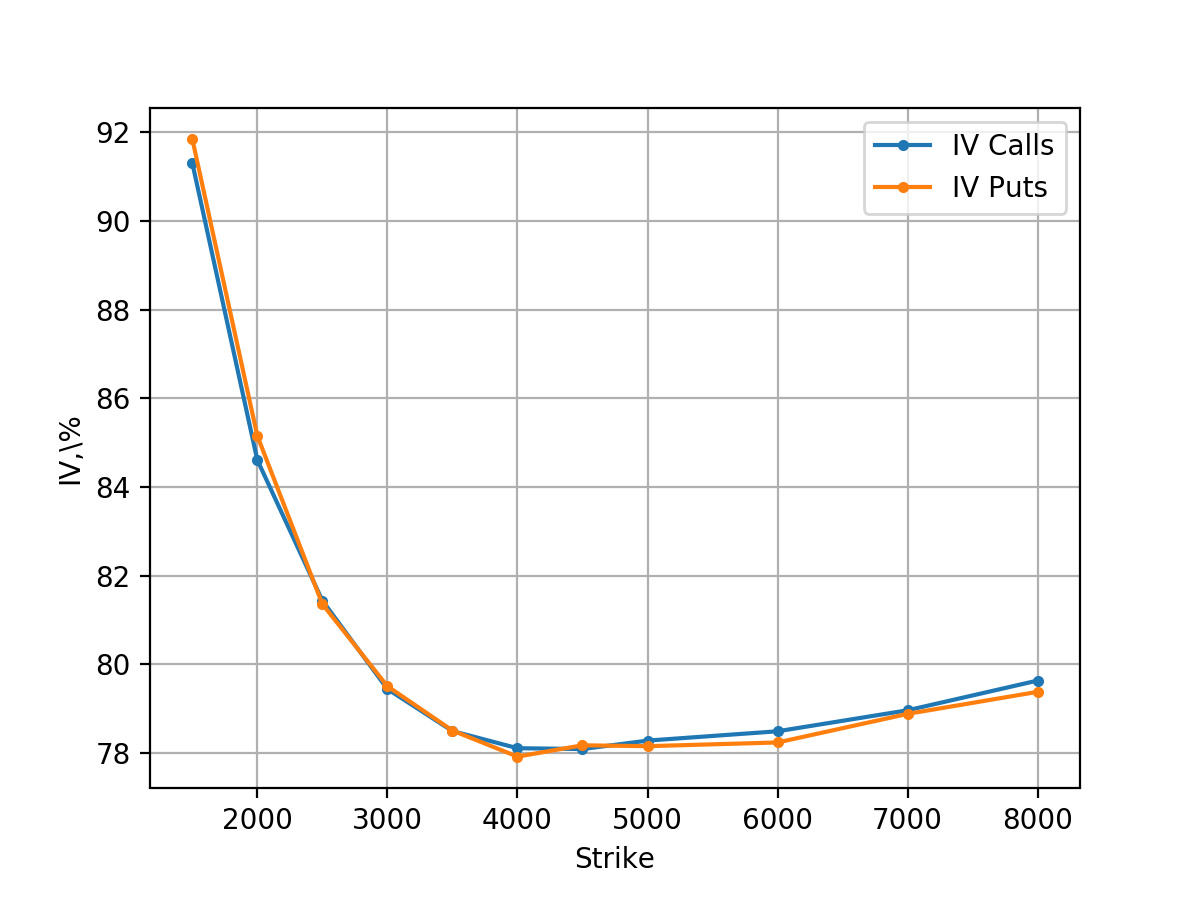}
\caption{Black-Scholes implied volatilities from options with maturities at 28-Jun-19 are used. Snapshot data are taken at 2018-Dec-11 04h:10m}
\label{fig:implied_volatilities}
\end{figure}

Black-Scholes implied log-normal volatilities replicate the well-known 'volatility smile' \cite{derman}. As an example, see Fig.~\ref{fig:implied_volatilities}.

'Volatility smile' is an artifact of Black-Scholes model and makes option pricing a very complicated technology. In order to replicate market prices, we have to build a complex theoretical structure, like interpolations of volatility surfaces or capture of correlated dynamics between volatilities and returns on forward prices. Use of "Logistic distribution" makes modeling significantly simpler.

\subsection{One day evolution of IPD}
\label{sec:one_day_ipd}

\begin{figure}[h]
\centering
\includegraphics[width=0.8\textwidth]{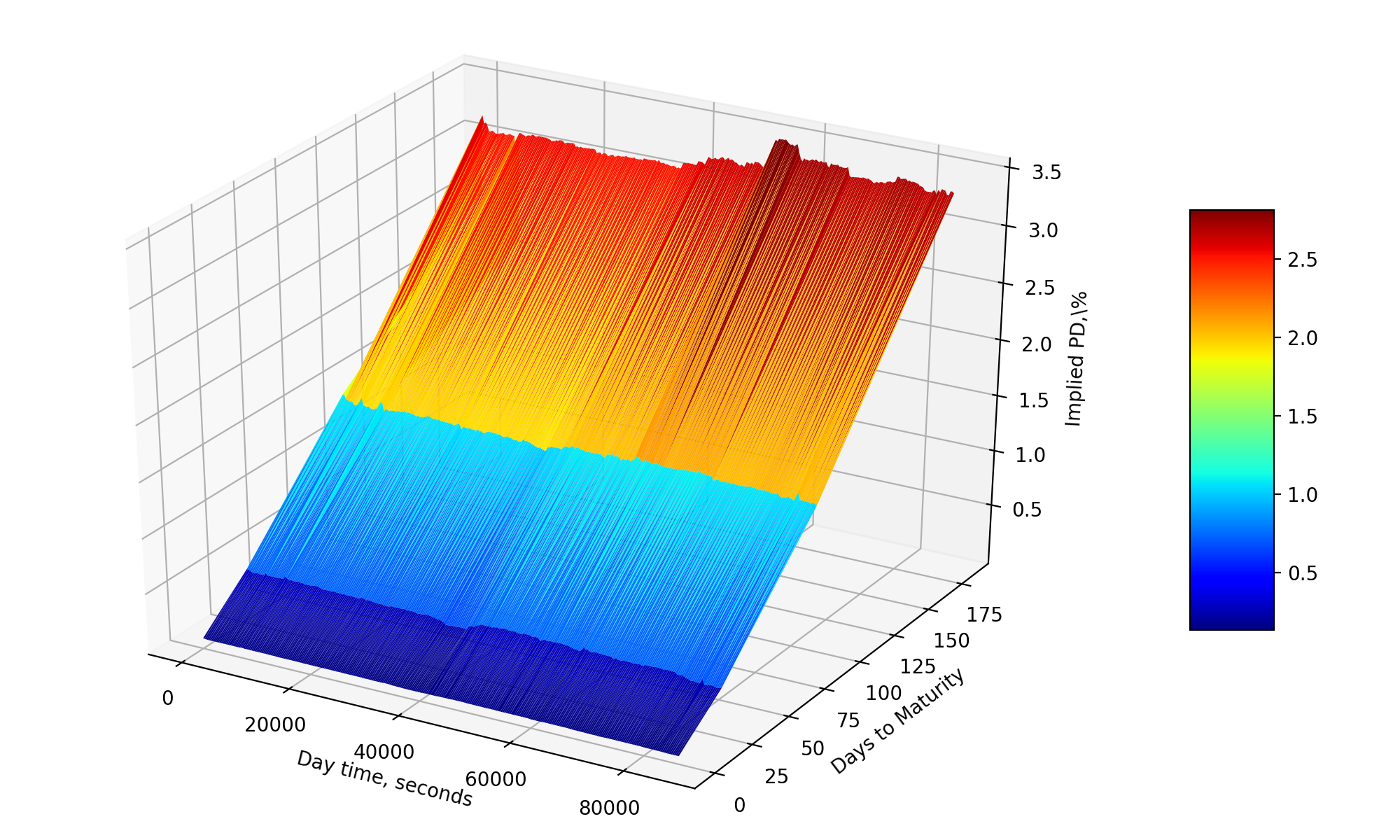}
\caption{283 snapshots at 5 minutes intervals from 11-Dec-2018 are analyzed. Implied PD (vertical $Z$-axis) is presented as function of Daytime in seconds ($X$) and Number of days to maturity ($Y$)}
\label{fig:one_day_ipd}
\end{figure}

Analysis of dataset from 11-Dec-2018 allows to see stability of developed procedures and certain correlations. Evolution of Implied PD's as function of day time and days to maturity is shown in Fig.~\ref{fig:one_day_ipd}

\subsection{Correlations between parameters}
\label{sec:parameter_correlations}

\begin{figure}[h]
\centering
\includegraphics[width=1.1\textwidth]{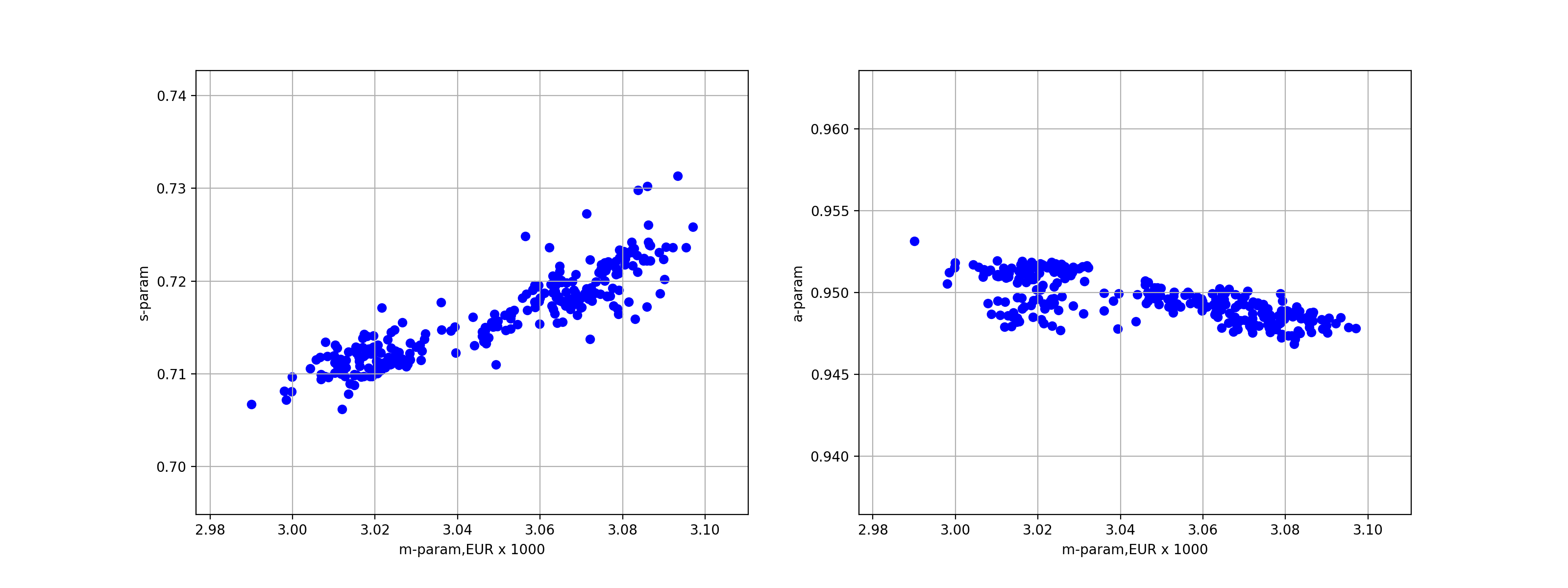}
\caption{Scatter plots of parameters. Left: $s$- vs $m$-parameters and Right: $a$- vs $m$-parameters. Data for options with maturity in March from 11-Dec-2018 are used.}
\label{fig:param_corr}
\end{figure}

Correlations between parameters are shown in Fig.~\ref{fig:param_corr} for completeness of results. $s$-parameter moves together with $m$-parameter, where changes correlate at 68\%, a-parameter is almost constant as function of $m$-parameter. Ideally, we would want to fix m-parameter equal to $ATM$ and $a$-parameter equal to 1 (because of definition of probability) leaving $s$-parameter as the only one free. 

Single parameter model is not illustrated here.
 
\subsection{Control of distributions}
\label{sec:controls}

\subsubsection{Comparison of model implied p.d.f. with future distribution}
\label{sec:comparison_pdf}

\begin{figure}[h]
\centering
\includegraphics[width=1.1\textwidth]{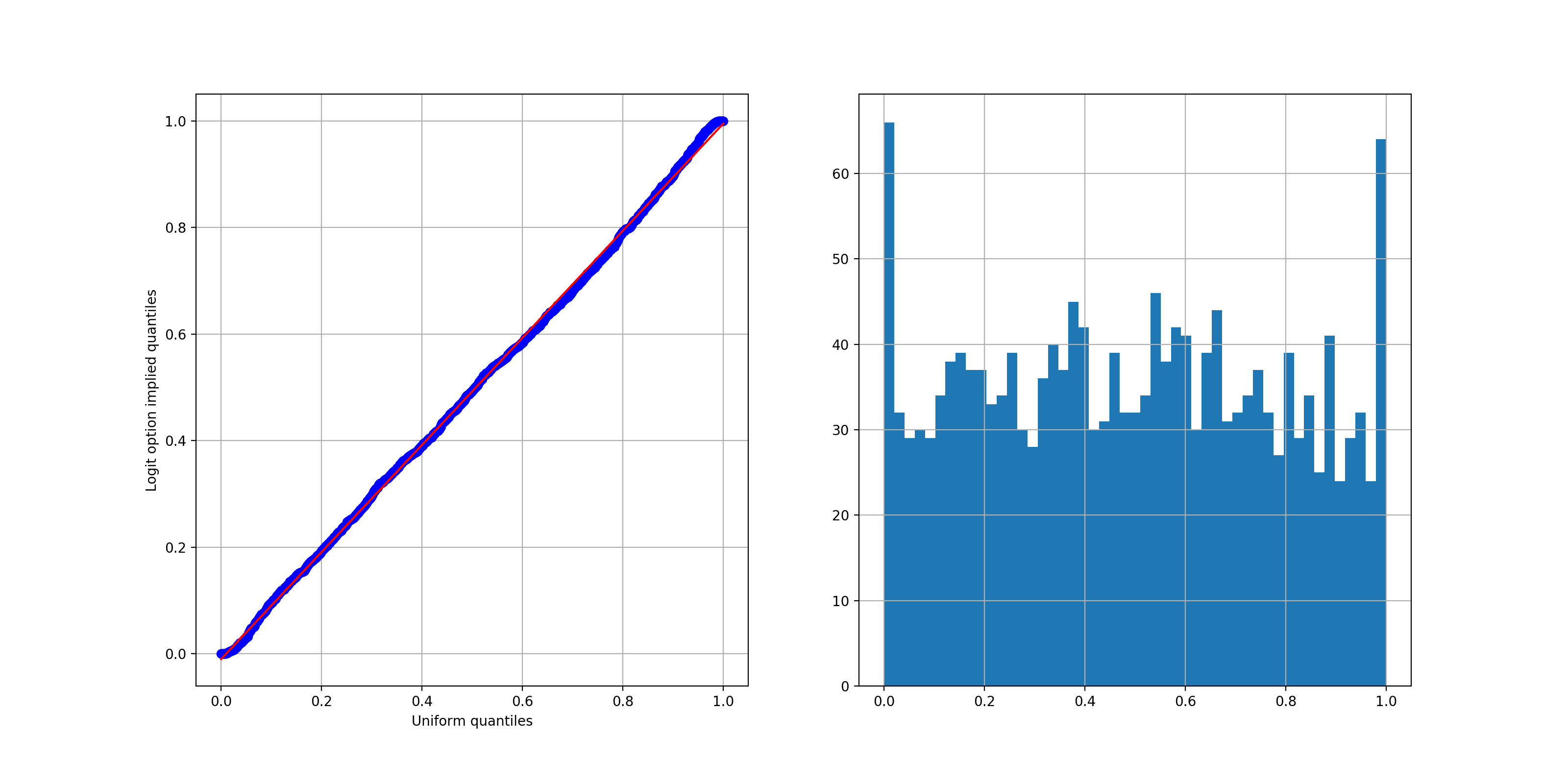}
\caption{Left: QQ-plot of quantiles of probabilities of historical returns extracted from CDF Implied from options (Y-axis) vs Uniform quantiles. Right: Option prices with Maturity of Jan-25-2019 are used. 1747 snapshots from 08-Dec-2018 to 18-Dec-2018 are analyzed.}
\label{fig:implied_quantiles_jan_norm}
\end{figure}

In order to validate interated sigmoid, $IS(x)$, as function to fit option prices we can compare implied p.d.f. with forward distribution of returns. For this purpose, for each moment in time $CDF(x)$ is implied from option prices and compared with forward return: 
\begin{equation}
r_{FWD}=ATM_{t}-ATM_{t-1}
\end{equation}
, where $ATM_t$ is implied from put-call parity and used as a proxy of futures. 5-minute returns are used. Implied $PDF$ is scaled down to 5-minute horizon from Maturity horizon by factor equal to root square of time, $\sqrt{(T-t)/5min}$. If return follows predicted distribution, then we can write:
\begin{equation} \label{eq:pdf_forward}
PDF_{5m}(x) \sim PDF(x, m, s) \cdot \sqrt{\frac{T-t}{5min}}
\end{equation}
Therefore, 
\begin{equation} \label{eq:pdf_forward}
CDF_{5m}^{-1} (r_{FWD}) \sim U(.)
\end{equation}
i.e. returns transformed with inverted implied CDF should follow uniform distributuon, U(.).The result is shown below in Fig.~\ref{fig:implied_quantiles_jan_jan}. Data seem to not follow the prediction given by options. By inroducing additional scaling factor to option implied distribution of $\eta=0.728$ we find that both forward returns and implied PDF follow similar shape, see Fig.~\ref{fig:implied_quantiles_jan_norm}. Kolmogorov-Smirnov p-value of comparison of $CDF_{5m}^{-1} (r_{FWD})$ to uniform distribution is 0.39. Scale $\eta<1$  indicates that option impied PDF is wider than the ‘real’ distribution of returns. It is a well known fact for option traders who overprice their options with respect to historical (realised) volatility to cover volatility of volatility risk. 


\begin{figure}[h]
\centering
\includegraphics[width=1.1\textwidth]{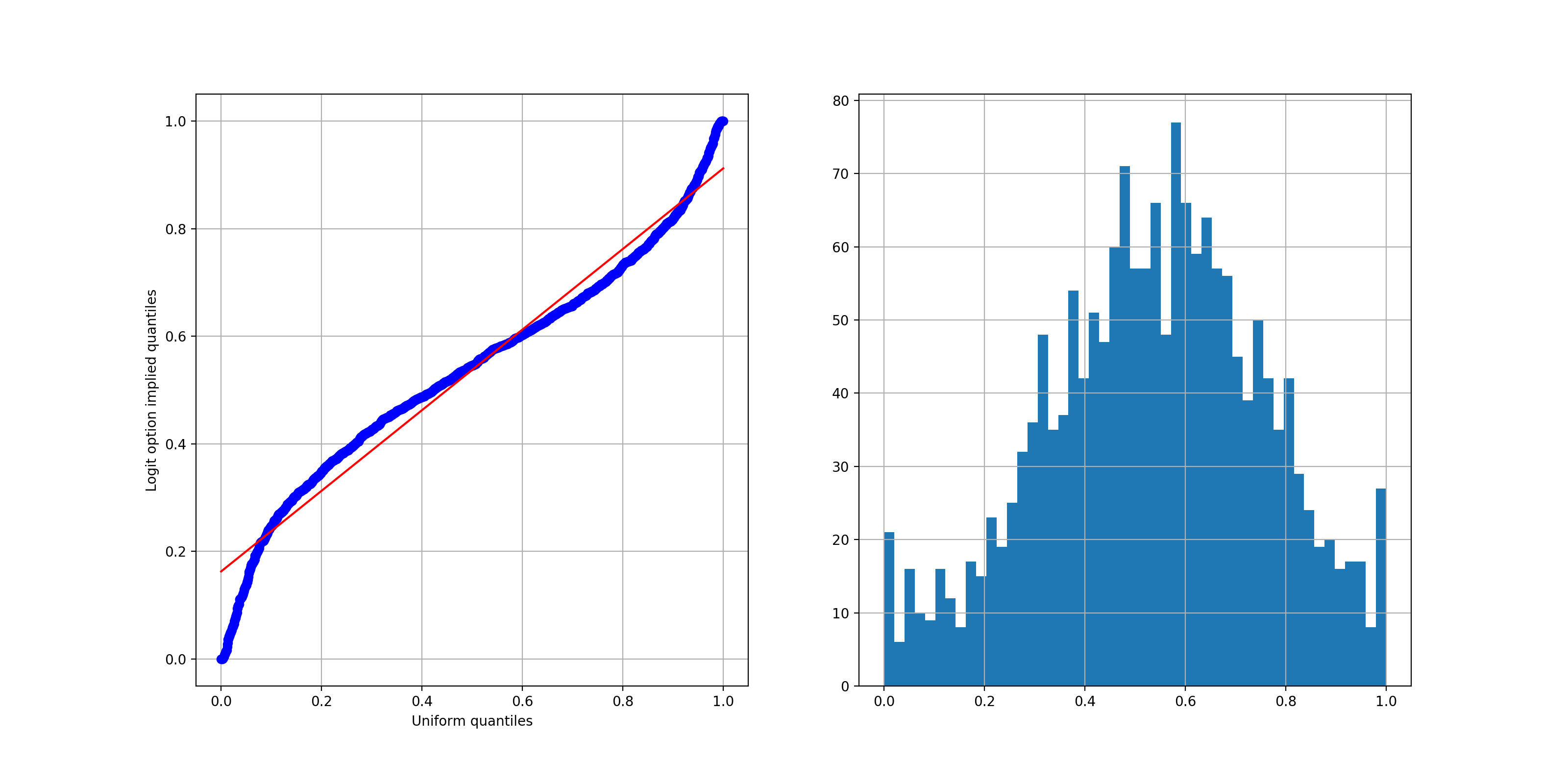}
\caption{Left: QQ-plot of quantiles of probabilities of historical returns extracted from CDF Implied from options (Y-axis) vs Uniform quantiles. Right: Option prices with Maturity of Jan-25-2019 are used. 1747	 snapshots from 08-Dec-2018 to 18-Dec-2018 are analyzed}
\label{fig:implied_quantiles_jan_jan}
\end{figure}

\section{Logistic distribution as distribution of underlying process}
\label{sec:logistic_as_process}

\subsection{New Parameters as risk factors}
\label{sec:new_parameters}

Parameters of presented logistic distribution, $m$, $s$, $a$, can be considered as new risk factors. If so, they can be used in hedging risks and risk management framework. In this case, $m$ can be interpreted as forward price (or ATM) level. $s$ can be used as replacement of volatility. When $m$ and $a$ are fixed as described above in section~\ref{sec:fit_prices_to_is} the model still provides satisfactory results. Since sigmoid and its 'siblings' are all expressed in closed form, the derivatives with respect to these parameters can be used to manage option positions more accurately.

However, the above proposal should be used with care, because only consistent pricing framework~\footnote{Black-Scholes framework is 'consistent' in the sense that all products including exotics can be priced and hence hedged with log-normal process of underlyiing which is shared among different products.} can make these parameters the real new risk factors. Otherwise they can be used only on the empirical basis. 

\subsection{Dynamics}
\label{sec:dynamics}

Remember, that the pricing framework based on Black-Scholes model allows non-arbitrage cross calibration of wide range of vanilla and exotic products with uniform interpretation of implied volatilities. Since logistic distribution is not a stable distribution, the creation of consistent pricing based on that seems to be very difficult at this point if not possible at all. Also, under Central Limit Theorem the distribution of returns simulated as $dr=\mu \cdot dt+dLD$, where $dLD \sim Logistic(x)$, will converge to the Normal distribution. This ‘problem’ remains unsolved. 
So far the the only way to make logistic distribution stable again is to combine it with geometric distribution as it is proposed in [6]. We do not see how this operation can be easily connected with trading mechanisms.

\subsection{Other asset classes}
\label{sec:other_assets}

The same logistic distribution was briefly applied to american equity options (e.g. Microsoft, MSFT traded at NASDAQ). The resulting fit was good with similar quality. Interest rate Swaptions (IR options on swaps as underlying) were also investigated with this method. The result is also satisfactory upto maturities and terms of underlying swaps of less than 20 years. From this we would like to cautiously conclude that this finding is universal. More analysis has to be performed to confirm this statement.

\section{Conclusion}
\label{sec:conclusion}

Analysis of BTC option prices is presented. We propose new method to describe option prices with integral sigmoid, which is CDS of logistic distribution. The analysis demonstrates great stability and ability to interpret parameters of this model. This parameterisation employs only 3 parameters. Model with only one free parameter is also satisfactory. Option implied PDF is compared to forward price movement and perfect agreement is found. It is still a question to answer whether this parameterization is due to 'pure luck' or it has some understandable dynamics, which can be used as an alternative to Black-Scholes log-normal model. Put-call parity for inverse options and futures traded within BTC exchanges is derived. 

\section{Acknowledgments}

This work is self-funded and is done as a part of service advertised at \textit{UXTA.io}. Although all ideas and analysis presented in this paper are original, the author would like to thank Oleg Nedbaylo for his support and collaboration in the development of algo-trading platform. This idea came as a direct conqsequence of this partnership. Author also would like to express the gratitude to Stefan Boor (Finmetrica Ltd) for proof reading and comments on the article. I express many thanks to Cyril Shmidt (Abn-AMRO) who also challendged some derivations.


\begin{thebibliography}{99}
\bibitem{bitcoin.org} list of bitcoin exchanges. at \url{https://bitcoin.org/en/exchanges#international}
\bibitem{coinmarketcap.com} list of bitcoin exchanges. at \url{https://coinmarketcap.com/exchanges/volume/24-hour/}
\bibitem{ABragin} "Inverse Futures in Bitcoin Economy", SSRN-2713755, 2015
\bibitem{breeden} Breeden, D.; Litzenberger, R.. "Prices of State-contingent Claims Implicit in Option Prices", Journal of Business, 51, pp. 621-651, 1978.
\bibitem{derman} Derman, E.; I, Kani.. "The Volatility Smile and Its Implied Tree". RISK, issue 7-2, 1994, pp. 139-145, pp. 32-39
\end{thebibliography}
\end{document}